\documentclass[aps,prd,amsmath,amssymb,superscriptaddress,altaffillsymbol, preprintnumbers,preprint,nofootinbib,a4paper]{revtex4}
\pdfoutput=1
\usepackage{amsthm}
\usepackage{graphicx}
	\graphicspath{{./figure/}}
\usepackage{color}
\usepackage{dcolumn}
\usepackage{bm}
\usepackage{bbm}
\usepackage{subfigure}
\usepackage{pxfonts}
\usepackage{slashed} 

\usepackage{epsfig}

\usepackage[ margin=5pt, font=normalsize,labelfont=bf,justification=raggedright]{caption}

\usepackage{hyperref}
\definecolor{rossoCP3}{cmyk}{0,.88,.77,.40}
\definecolor{verdeCP3}{rgb}{0.09765625, 0.57421875, 0.1015625}
\definecolor{bluCP3}{rgb}{0, 0.23, 0.67}
\hypersetup{colorlinks, bookmarksopen, bookmarksnumbered,citecolor=verdeCP3, linkcolor=bluCP3, pdfstartview=FitH, urlcolor=rossoCP3}
\def\lsim{\mathrel{\rlap{\lower4pt\hbox{\hskip1pt$\sim$}}
    \raise1pt\hbox{$<$}}}                
\def\gsim{\mathrel{\rlap{\lower4pt\hbox{\hskip1pt$\sim$}}
    \raise1pt\hbox{$>$}}}                

\newcommand{\ea}[1]{
\begin{align}
#1
\end{align}
}

\newcommand{\be}{\begin{eqnarray}}
\newcommand{\ee}{\end{eqnarray}}

\begin{document}
\title{\Large  \color{rossoCP3} ~~\\
 \color{black} $S$\color{rossoCP3} table \color{black} $E \chi$\color{rossoCP3} tensions with(out) Gravity}
\author{Oleg Antipin}
\email{antipin@cp3-origins.net} 
\author{Jens Krog}
\email{krog@cp3-origins.net} 
\affiliation{{\color{rossoCP3} CP$^{3}$-Origins} \& Danish Institute for Advanced Study {\color{rossoCP3} DIAS}, University of Southern Denmark, Campusvej 55, DK-5230 Odense M, Denmark}
\author{Matin Mojaza}
\email{mojaza@cp3-origins.net} 
\author{Francesco  Sannino}
\email{sannino@cp3-origins.net} 
\affiliation{{\color{rossoCP3} CP$^{3}$-Origins} \& Danish Institute for Advanced Study {\color{rossoCP3} DIAS}, University of Southern Denmark, Campusvej 55, DK-5230 Odense M, Denmark}
\newpage
\begin{abstract}
We investigate the vacuum stability as well as the gravitational corrections in extensions of the Standard Model featuring a new complex scalar, and two Dirac fermions for different choices of the hypercharge of the scalar and one of the two fermions. The neutral fermion acquires loop-induced magnetic interactions with the Standard Model and could be identified with a dark matter candidate. 
To the lowest order in perturbation theory we show that these extensions can save the electroweak vacuum from being metastable.  We then add the gravitational corrections to the different beta functions and discover that the models can be compatible with the asymptotically safe gravity scenario at the price of a heavier Higgs and lighter top mass.   
\\[.1cm]
{\footnotesize  \it Preprint: CP$^3$-Origins-2013-44, DIAS-2013-44.}
 \end{abstract}

\maketitle

\section{{$S\overline{E}\chi \lowercase{y}$} extensions of the Standard Model}

With the recent discovery \cite{:2012gk,:2012gu} of a new resonance with properties similar to the Standard Model (SM) Higgs particle, one of the greater concerns when studying the ultraviolet (UV) behavior of the SM has become whether or not its ground state is stable all the way to the Planck scale \cite{Degrassi:2012ry}. If unstable this might indicate the need for new physics to occur to stabilize the theory. At the Planck scale and beyond unknown gravity corrections could ensure the stability of  the theory at those energies. Indeed, it was proposed in  \cite{Shaposhnikov:2009pv,Bezrukov:2012sa} that gravity can be directly responsible for a vanishing quartic Higgs coupling at the Planck scale. This UV constraint leads to the prediction for the Higgs mass of about $129\pm 6$ GeV. However with the recent higher loop refinements of the high energy behavior of the SM potential it is clear that the Higgs quartic coupling does not stay positive till the Planck scale. Instead the potential develops a second minimum at high field values, making the electroweak vacuum, at best, metastable   \cite{EliasMiro:2011aa,Degrassi:2012ry,Chetyrkin:2013wya,Antipin:2013sga}. 

These findings do not automatically imply that the SM holds an inconsistency by itself, but this behavior {is not in line with the asymptotically safe gravity scenario}. In addition, the SM is already known to be incomplete and
it is now particularly timely to investigate minimal extensions of the SM featuring new sectors, possibly relevant for the dark matter problem.
Here we investigate dark matter motivated extensions \cite{Dissauer:2012xa,Frandsen:2013bfa} of the SM, where dark matter is magnetically interacting. Here, for the first time we investigate their possible simultaneous ability to  save the electroweak vacuum from being metastable, provide a dark matter candidate, and their compatibility with the asymptotically safe gravity framework \cite{Shaposhnikov:2009pv,Bezrukov:2012sa}.

The new sector consists of a vectorlike heavy electron ($E$), a complex heavy scalar electron ($S$) and a SM singlet Dirac fermion ($\chi$). The associated renormalizable Lagrangian is 
\begin{eqnarray}
\label{Sexy}
    \mathcal{L}_{S\overline{E}\chi {\rm y}}& =& \mathcal{L}_{\text{SM}}+\bar{\chi}i\slashed{\partial}\chi - m_\chi\bar{\chi}\chi + \overline{E}i\slashed{D}E - m_E\overline{E}E - (S \overline{E}{\chi}  y_\chi + \text{h.c.} ) \nonumber \\
& +& D_\mu S^{\dagger}D^\mu S - m_S^2S^{\dagger}S-\lambda_{HS}H^{\dagger}HS^{\dagger}S - \lambda_S (S^{\dagger}S)^2\ ,\end{eqnarray}
where $H$ is the SM Higgs doublet and $D^\mu=\partial^\mu - i g_1 Q_D B^\mu$, with $g_1$ the hypercharge coupling and $Q_D$ denoting the hypercharge of $E$ and $S$.
We assume the new couplings $y_\chi, \lambda_{HS}$ and $ \lambda_S$ to be real and the bare mass squared of the $S$ field, $m_S^2$, to be positive  so that electroweak symmetry breaks via the Higgs doublet. 
The interactions among $\chi$, our potential magnetic dark matter candidate, and the SM fields occur via loop-induced processes involving the $S\bar{E} \chi y$-operator.
The scalar electron $S$ has properties reminiscent of a selectron except that it is vectorlike, and therefore only the scalar field $S$ feels the Higgs directly. This is true provided we do not mix the new electron with the SM leptons via generalized Yukawa interactions. Due to this property and the requirement of the renormalizability of the theory, the $S$ sector is a portal sector and can be probed directly using processes involving the Higgs.

The phenomenological signatures of this model were studied in Ref.~\cite{Dissauer:2012xa}, where it was constructed in the 
search for a theory that is able to alleviate the tension between the different direct-detection dark matter searches \cite{An:2010kc,Banks:2010eh,DelNobile:2012tx,Fitzpatrick:2010br}. 
%
This model, without the explicit mass parameters, was also recently considered as a \emph{perturbatively natural conformal} extension of the SM  \cite{Antipin:2013exa},
where electroweak symmetry breaking is generated via the Coleman-Weinberg mechanism without any quadratic divergences to the perturbative order considered. This scenario, in fact,  predicts the mass of $S$ to be around $m_S\approx 383$\  GeV, close to the benchmark value used in \cite{Dissauer:2012xa}.

A more detailed analysis of the dark matter properties and constraints of these theories appeared in \cite{Frandsen:2013bfa}. Here it was shown that the basic model is constrained dominantly by direct detection experiments and its parameter space can be nearly entirely covered by up-coming ton-scale direct detection experiments. It is clear that adding the vacuum stability analysis and the interplay with gravitational interactions allows us to get one step closer to a more complete extension of the SM.

The paper is structured as follows.  In section \ref{RG} we show that it is possible to achieve electroweak vacuum stability. Here we also determine the regions, in coupling space, of instability, metastability and stability of the theory. We study the gravitational transition in section \ref{UVcompletion} by adding at the expected gravitational corrections to the beta functions of the theory.  Due of the nature of these corrections the scalar couplings and their beta functions are  expected to vanish at the transition scale, in absence of the gravitational corrections. We show that it is possible to abide to these conditions, but that generally they tend to bring these extensions towards the region of metastability. We offer our conclusions in section \ref{conclusions}.

\section{{$S\overline{E}\chi \lowercase{y}$} RG analysis and Vacuum Stability }
\label{RG}

We will first investigate the SM vacuum stability and finiteness of the 
running couplings as functions of the renormalization group (RG) energy scale $\mu$. This will give us a better understanding of the UV behaviour of the 
SM under the influence of the dark sector.
The following set of couplings are relevant to consider: The gauge couplings $g_1,g_2,$ and $g_3$, associated to the $U(1),SU(2)$ and $SU(3)$ gauge symmetry respectively, as well as the top Yukawa coupling $y_t$,  the Yukawa coupling of the dark sector $y_{\chi}$, and the three quartic couplings $\lambda_H,\lambda_{HS}$, and $\lambda_S$.

Without gravitational corrections, i.e. in the low-energy region, their respective beta functions are given to one loop order by:
\begin{align}
\label{g1beta}
\beta_{g_1}&=\frac{1}{(4\pi)^2}\left(\frac{41}{6}+\frac{5}{3}Q_D^2\right)g_1^3 \quad,\quad
\beta_{g_2}=-\frac{19}{96\pi^2}g_2^3  \quad,\quad
\beta_{g_3}=-\frac{7}{(4\pi)^2}g_3^3 \ ,\\
\beta_{y_t}&=\frac{1}{(4\pi)^2}\left[ \frac{9}{2}y_t^3-(\frac{17}{12}g_1^2 +\frac{9}{4}g_2^2+8g_3^2)y_t  \right] \ ,\\
\label{yxbeta}
\beta_{y_{\chi}}&=\frac{1}{(4\pi)^2}3y_{\chi}(y_{\chi}^2-Q_D^2g_1^2) \ ,\\
\beta_{\lambda_{H}} &= \frac{1}{(4\pi)^2}\left[3 (4y_t^2-3g_2^2-g_1^2)\lambda_H -6y_t^4 +\tfrac{3}{8}[2g_2^4+(g_1^2+g_2^2)^2]+24\lambda_{H}^2+\lambda_{HS}^2\right] \ ,\\
\beta_{\lambda_{HS}} &= \frac{1}{(4\pi)^2}\left[ \frac{3}{2}(4y_t^2-3g_2^2-g_1^2+8\lambda_H)\lambda_{HS}+ (4y_{\chi}^2-6Q_D^2g_1^2+8\lambda_{S}+4\lambda_{HS})\lambda_{HS}+3Q_D^2g_1^4 \right] \ ,\\
\beta_{\lambda_{S}}&= \frac{1}{(4\pi)^2}\left[2\lambda_{HS}^2+6Q_D^4g_1^4-12Q_D^2g_1^2\lambda_{S} +20\lambda_{S}^2+8y_{\chi}^2\lambda_{S}-4y_{\chi}^4 \right] 
\label{lsbeta}\ ,
\end{align}
where $Q_D$ is the hypercharge of the $E$ and $S$ fields.
Perturbative couplings of the new sector are compatible with the phenomenological constraints presented in \cite{Dissauer:2012xa}, and therefore this set of beta functions can be applied around the Fermi scale%
\footnote{In order to do a precise RG analysis, the Weyl consistency conditions have to be respected. This implies that to investigate the one-loop evolution of the scalar quartic couplings, one must take into account the three loop beta functions for the gauge couplings and the two loop beta functions for the Yukawa couplings  \cite{Antipin:2013sga,Antipin:2013pya}. The goal of this work is to obtain a rough understanding of the RG flow, for which the one loop analysis in all couplings is sufficient.
To this end we must assume that the couplings of the extended sector remain small along the RG flow.}.
We will assume, that the DM candidate $\chi$ has a mass around $m_{\chi}\sim 10 $ GeV, and that the mass of the scalar and vector like electron have masses $m_S \sim m_E \sim 500 $ GeV.

Due to the decoupling theorem, the SM couplings will run as in the SM, until the mass scale $m_S$ of the new scalar $S$ (and electron $E$) is reached. To lowest order in perturbation theory, there is no threshold effects on the couplings at this scale, since the vacuum expectation value of $S$ is at the origin. Beyond the $m_S$ scale the running couplings are influenced by the new sector. 
In particular, at one loop, the beta function for the $U(1)$ gauge coupling, $g_1$, is modified since the new scalar $S$ appeaars in the loop corrections to the $g_1$ coupling, and the beta function for the Higgs self-coupling $\lambda_H$ receives corrections from the portal coupling, $\lambda_{HS}$.
Defining values of $\lambda_S$, $\lambda_{HS}$ and $y_{\chi}$ at the $m_S$ scale as well as choosing a value for the hypercharge $Q_D$, will then uniquely dictate the evolution of the theory, at least until gravitational corrections should be taken into account. 

{ 
In order to constrain the parameter space of the theory, we will look for fixed point structures in the new sector.  Upon inspection of \eqref{g1beta} and \eqref{yxbeta}, we find that the ratio $\frac{y_{\chi}}{g_1}$ has an \emph{approximate} IR fixed point, which reads:
\ea{
\label{y_chi-g1-ratio}
\left. r \equiv \frac{y_{\chi}}{g_1}\right|_{IR}=\sqrt{\tfrac{41}{18}+\tfrac{14}{9}Q_D^2} \ .
} 
Assuming that $y_\chi$ reaches small values in the IR of order $g_1$  (i.e. at the $m_S$ scale), we can expect that the ratio of these couplings in the IR is close to this value. 
We impose this assumption in our analysis to determine $y_{\chi}(m_S)$ from $g_1(m_S)$.

Furthermore, the beta function of $\lambda_S$ in Eq.~\eqref{lsbeta} is, except a pure $\lambda_{HS}^2$ term, only dependent on the couplings $\lambda_S$ and $y_{\chi}$, when replacing $g_1$ with the above constraint.  
There is a fixed point in $\lambda_S$, which we can express in terms of the ratio $\frac{\lambda_S}{y_{\chi}^2}$ after replacing $g_1$, reminiscent of the quartic-gauge approximate fixed point of the SM. If $\lambda_{HS}$ at low energies is of the order of $y_{\chi}^2$ and we keep $Q_D<3$, then after having defined $\kappa = (Q_D/r)^2$ we have:
\ea{
\label{lambda_S-y_chi-ratio}
\left . \frac{\lambda_S}{y_{\chi}^2}\right|_{IR} = \frac{3 \kappa -1}{20} + \frac{\sqrt{81-6\kappa - 111\kappa^2}}{20} + \mathcal{O}(\lambda_{HS}) \ .
}
The leading estimate varies between 0.42 and 0.32 for any value of $Q_D$.
Thus, as before, by assuming $\lambda_S$ to reach small values in the IR of order $y_\chi^2$, we can expect this ratio to be fulfilled at the $m_S$ scale.
We impose also this assumption to simplify the RG analysis.
}

 The remaining parameters of the model that needs to be specified are $Q_D$ and $\lambda_{HS}(m_S)$. In the analysis below we require absolute stability of the electroweak vacuum at least up to the Planck scale. 
Thus, for a specific value of $Q_D$ we analyze the running couplings and the effective potential, and set a lower limit on $\lambda_{HS}(m_S)$ that ensures absolute stability.

\begin{figure}[bt]
\includegraphics[width=.48\textwidth]{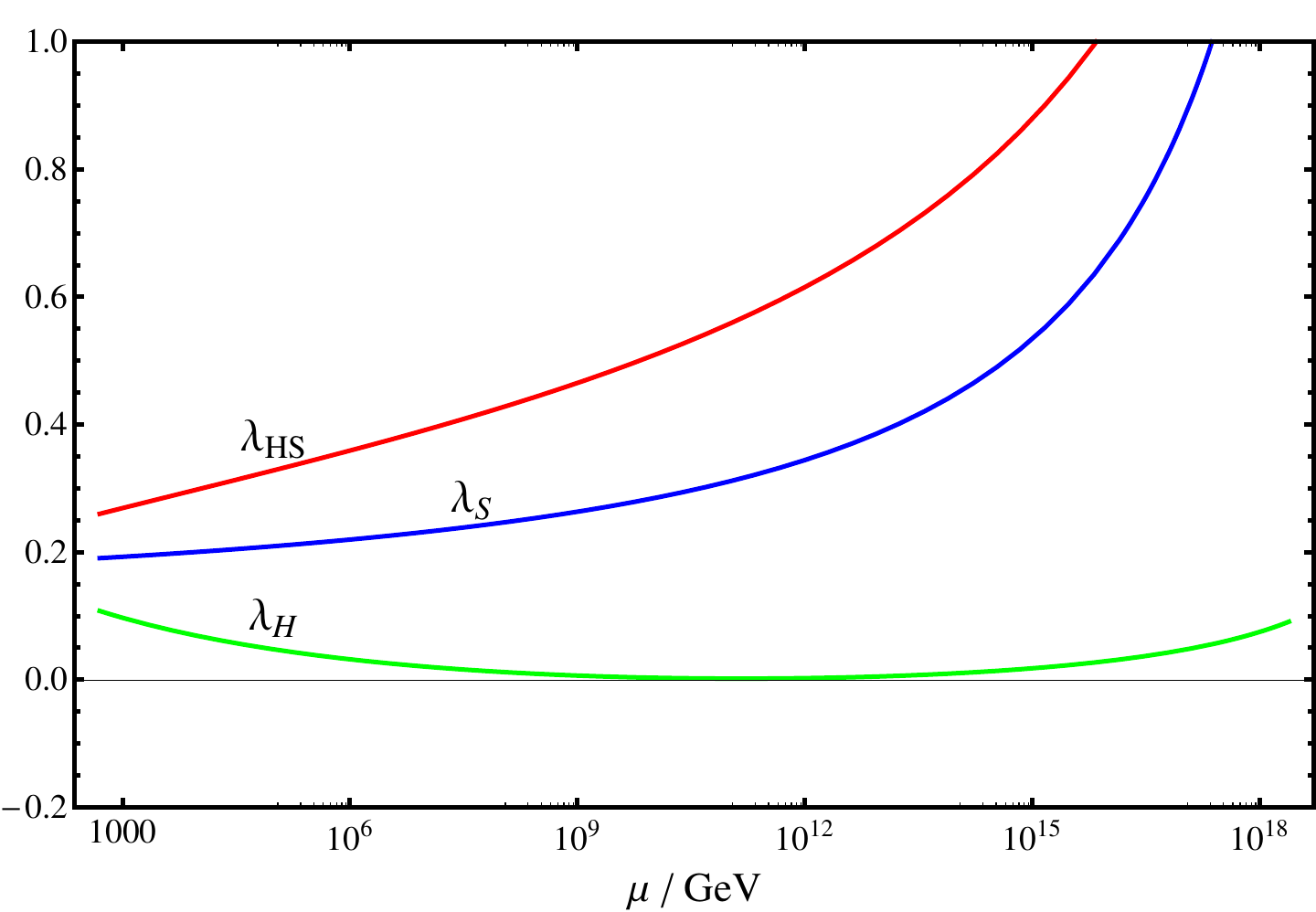}
\includegraphics[width=.48\textwidth]{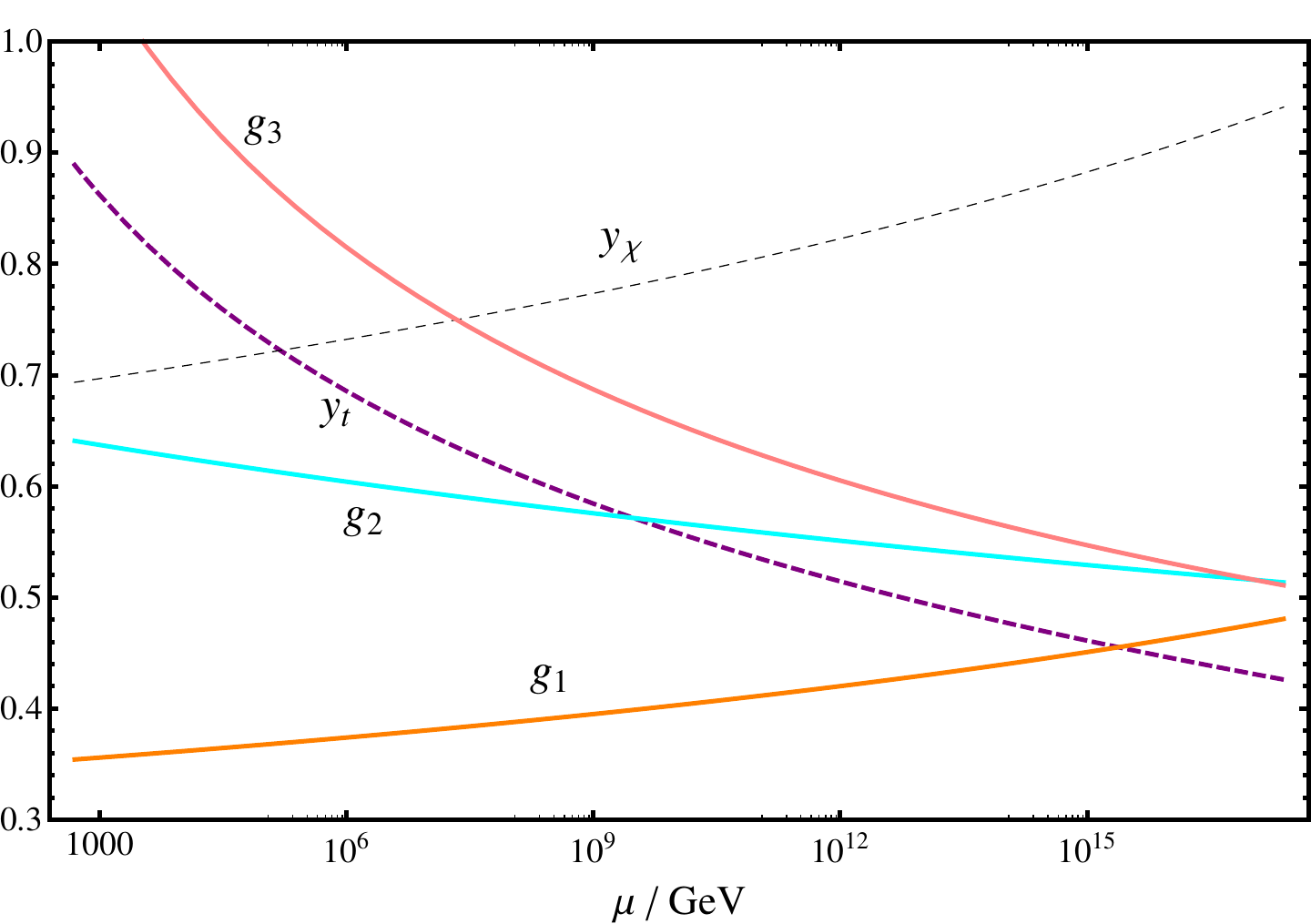}
\caption{RG evolution of the couplings, where $Q_D=1$, $m_S=500$ GeV, $\lambda_{HS}(m_S)=0.26$, $y_{\chi}(m_S)=0.69$ and $\lambda_S(m_S)=0.19$. The Higgs self coupling is stabilized due to the portal coupling $\lambda_{HS}$, which is here at its lower bound to ensure stability.}
\label{fig:IR_QD1}
\end{figure}

\begin{itemize}
\item[For $Q_D=1$] we deduce from Eq.~\eqref{y_chi-g1-ratio} and \eqref{lambda_S-y_chi-ratio} that $y_{\chi}(m_S)\approx 0.69$, and $\lambda_S(m_S) \approx 0.20$ up to corrections from $\lambda_{HS}$.
The smallest value for $\lambda_{HS}$ that ensures stability of the electroweak vacuum is $\lambda_{HS}(m_S) \gsim 0.26$. A nonzero value of $\lambda_{HS}$ leads, in the full analysis, to a slightly smaller $\lambda_S(m_S) \lsim 0.19$. The running couplings are shown in Figure \ref{fig:IR_QD1}. 
The quartic couplings run to large values around the Planck scale.
Thus the lower boundary on the coupling $\lambda_{HS}$ corresponds to
a Landau pole close to the Planck scale.  For higher values of $\lambda_{HS}$ the Landau pole is shifted toward lower energy scales.

\item[For $Q_D=2$] 
we have $y_\chi(m_S) \approx 1.03$ and the lower boundary
on $\lambda_{HS}$ is slightly lowered; \mbox{$\lambda_{HS}(m_S) \gsim 0.2$} with $\lambda_S(m_S) \lsim 0.41$. The Landau pole, however, is also lowered to around the value $10^{15}$~GeV. For even higher values of $Q_D$ this trend continues and the value of $y_\chi(m_S)$ quickly becomes non-perturbative.

\item[For $Q_D \sim 0$] 
i.e. for millicharged dark scalar and electron, the trend goes in the opposite direction; the lower bound becomes $\lambda_{HS} \gsim 0.28$ with $y_\chi(m_S) \approx 0.53$ and $\lambda_S(m_S) \lsim 0.11$ and the Landau pole moves beyond the Planck scale.
\end{itemize}

For the case $Q_D = 1$ 
we make an elaborate study of the Higgs potential stability in the phase space of couplings.
The electroweak vacuum is not stable if the 
Higgs self-coupling $\lambda_H$ runs to negative values.
We can, however, distinguish metastability from
instability. 
This is done by considering the probability of tunneling to the 
true  vacuum during the evolution of the Universe.
If the probability is bigger than some value $p$, we say 
that the electroweak vacuum is unstable. Otherwise it is
metastable, and thus physical (see Refs.~\cite{Coleman:1977py,Espinosa:1995se,Isidori:2001bm} for details).
In our study, we choose the value $p = 0.1$, which means
that most of the space (more precisely $e^{-p} \sim 90 \%$) is in the metastable phase
at current times.

\begin{figure}[bt]
\includegraphics[width=.7\textwidth]{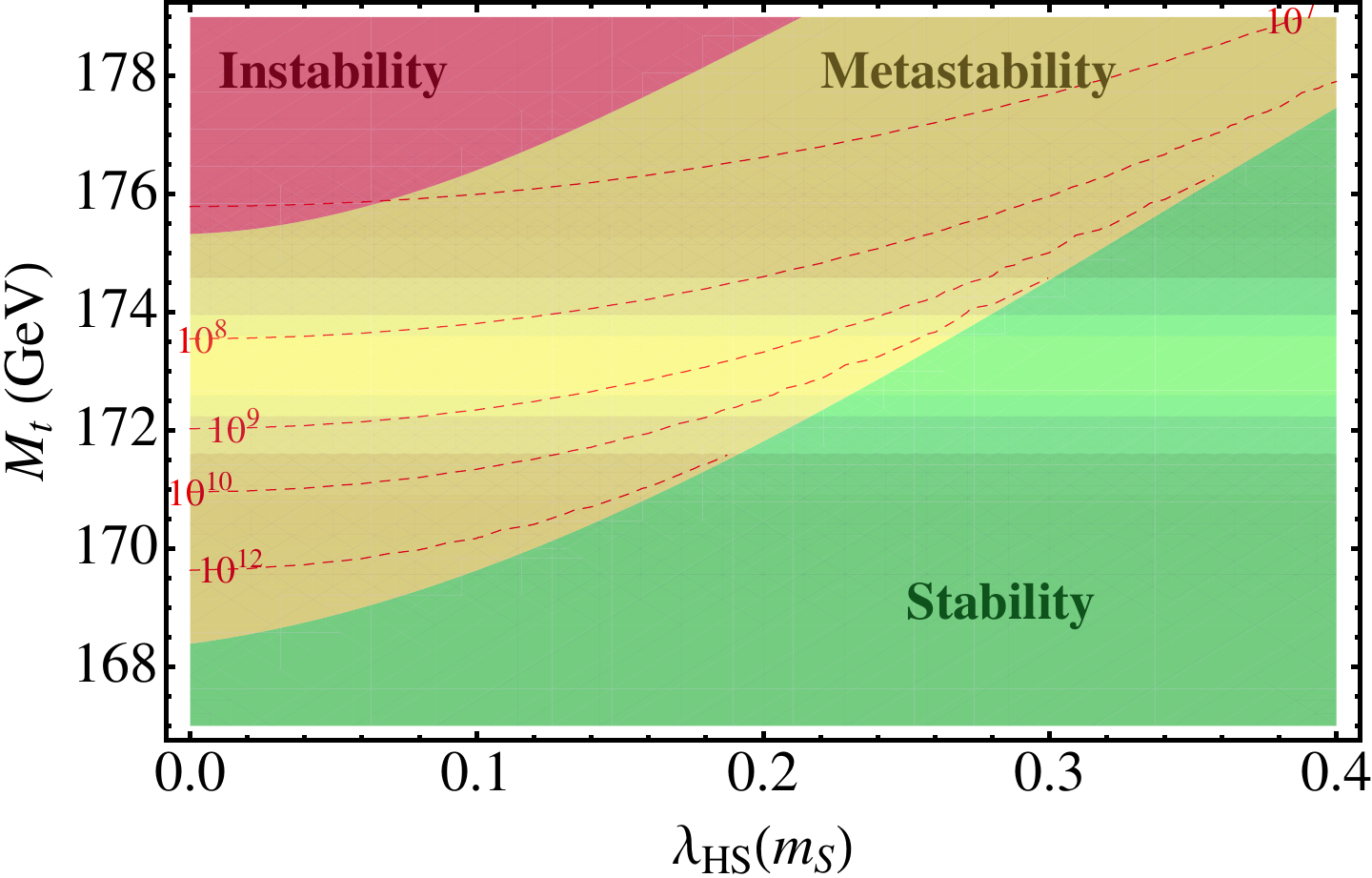}
\caption{The Higgs potential stability as a function of the top-quark mass $M_t$ and
$\lambda_{HS}(m_S)$. The shadings show the normal distribution of the top-mass with mean value $173.1$ GeV and standard deviation $\sigma = 0.9$ GeV, as given by the Particle Data Group \cite{Beringer:1900zz}. All other parameters were fixed to their central experimental values, in particular $m_H = 125.9$ GeV \cite{Beringer:1900zz}. The dashed contours indicate the scale (in GeV) where $\lambda_H = 0$.}
\label{sexymoney}
\end{figure}

In  Fig.~\ref{sexymoney} we show the results
of this analysis as a function of the top-quark mass $M_t$ and
$\lambda_{HS}(m_S)$, where we kept fixed
all other parameters fixed to their central experimental value, in particular $m_H = 125.9$ GeV, as given by the Particle Data Group \cite{Beringer:1900zz}. 
Varying $m_H$ within the experimental uncertainty does not generate any numerically significant difference in the figure.

So far we concentrated on the stability analysis. By combining it with the request of a viable dark matter candidate \cite{Frandsen:2013bfa}, typically needing large values of $\lambda_{HS}$ and $y_\chi$ at the electroweak scale, we conclude that the model is able to solve the dark matter problem while remaining stable. However, the scalar couplings are expected to generate a Landau pole before reaching the Planck scale.

\section{Crossing the gravity scale}\label{UVcompletion}
Near the four-dimensional Planck scale we can no longer ignore the gravitational corrections. Currently there is no universal consensus on how quantum gravitational corrections have to be dealt with. To progress here we will make use of the intriguing scenario  according to which quantum gravity becomes asymptotically safe, and therefore nonperturbatively renormalizable due to the occurrence of a strongly coupled UV fixed point  \cite{Weinberg}. The literature on the subject is vast and we refer to \cite{Litim:2008tt} for a review. To determine the gravitational corrections we follow \cite{Percacci:2003jz}.

Furthermore, the authors in \cite{Shaposhnikov:2009pv,Bezrukov:2012sa} noticed an intriguing feature of the SM when assuming the electroweak vacuum to be the true vacuum; i.e. they showed that a lower bound on the Higgs mass consistent with asymptotic safe gravity is $129 \pm 6$~GeV. These results seem to imply that the electroweak scale is somehow determined by Planck scale physics.

Here we test whether this picture survives, when including the effects of the candidate dark $S\bar{E}X y_\chi$ sector. 
Denoting collectively the set of dimensionless couplings by $x_i$ it follows from pure dimensional grounds that 
 the gravitational contribution, $\beta_i^{\rm grav}$, to the beta function of $x_i$  reads:
\begin{equation}
\label{betaigrav}
\beta_{x_i}^{grav} = \frac{a_i}{8\pi}\frac{\mu^2}{M_P^2(\mu)}x_i \ ,
\end{equation}
where the Planck scale, $M_P(\mu)$, is a dynamical quantity and
scales due to asymptotic safety as \cite{Shaposhnikov:2009pv}:
\begin{equation}
\label{mpk}
M_P^2(\mu)= M_P^2+2\xi_0 \mu^2
\end{equation}
where $M_P = (8\pi G_N)^{-1/2} =2.4\times  10^{18}\text{ GeV}$ is 
the usual (low energy) Planck mass. The parameter $\xi_0$ is a model and scheme dependent number. Its exact value is not important for this work and we fix its numerical value to $\xi_0 = 0.024$ based on numerical studies in certain (FRGE) gravity models \cite{Reuter:1996cp,Percacci:2003jz,Narain:2009fy}. 
Also the coefficients $a_i$ are scheme and model dependent and are furthermore dynamical. For our study only their sign near the Planck scale will be important.
The full one loop beta functions for the couplings $x_i$ thus read:
\begin{equation}
\label{betaigravII}
\mu\frac{dx_i}{d\mu}=\beta_{x_i}+\frac{a_i}{8\pi}\frac{\mu^2}{M_P^2(\mu)}x_i  \ . 
\end{equation}

The corrections to the beta functions from gravity are negligible until we reach $\mu^2 \sim \frac{M_P^2}{2\xi_0}$. 
If the couplings stay perturbative in the high energy regime, they are well described by Eq.~\eqref{betaigravII} with $a_i$ constant. For $\mu^2 > \frac{M_P^2}{2\xi_0}$ the gravitational
corrections become increasingly important.
In particular, for $a_i < 0$ the couplings will run towards zero in the UV, making them all asymptotically free. In Ref.~\cite{Shaposhnikov:2009pv} it was argued%
\footnote{
The argument for $a_{gauge}<0$ follows from explicit calculations in~ \cite{Robinson:2005fj,
Folkerts:2011jz}. The argument for $a_{Yukawa}< 0$ follows by negation, since positive values lead to trivial IR fixed points with $y_{t,IR}=y_{\chi,IR}=0$ (where IR is now the Planck scale as seen from the asymptotically safe UV fixed point), up to contributions from the gauge sector, which are not able to explain the large value of the top Yukawa coupling. { Negative values of $a_{Yukawa}$ are moreover supported by explicity computations~\cite{Rodigast:2009zj}.}}
 that $a_i$ for the gauge and Yukawa couplings are indeed expected to be negative, while explicit computations for $\beta_{\lambda_H}^{grav}$ yields $a_\lambda>0$ \cite{Percacci:2003jz,Narain:2009fy}.{ {One should note that 
 different results have been obtained in the literature \cite{Rodigast:2009zj}, 
 so a positive $a_\lambda$ is at this point an explicit assumption. }}
Due to the universal nature of gravitational interactions, these arguments apply equally to the couplings of the extended sector and therefore we assume the sign of the gravitational coefficient $a_i$ of each type of coupling to be: $a_{gauge} <0$, $a_{Yukawa} <0$ and $a_{quartic} > 0$. Thus also $y_\chi$ becomes asymptotically free beyond the Planck scale, while the quartic couplings $\lambda_H$ and $\lambda_S$ must both be positive or zero at the Planck scale to ensure
that the potential stays bounded from below beyond the Planck scale.

\begin{figure}[bt]
\includegraphics[width=.48\textwidth]{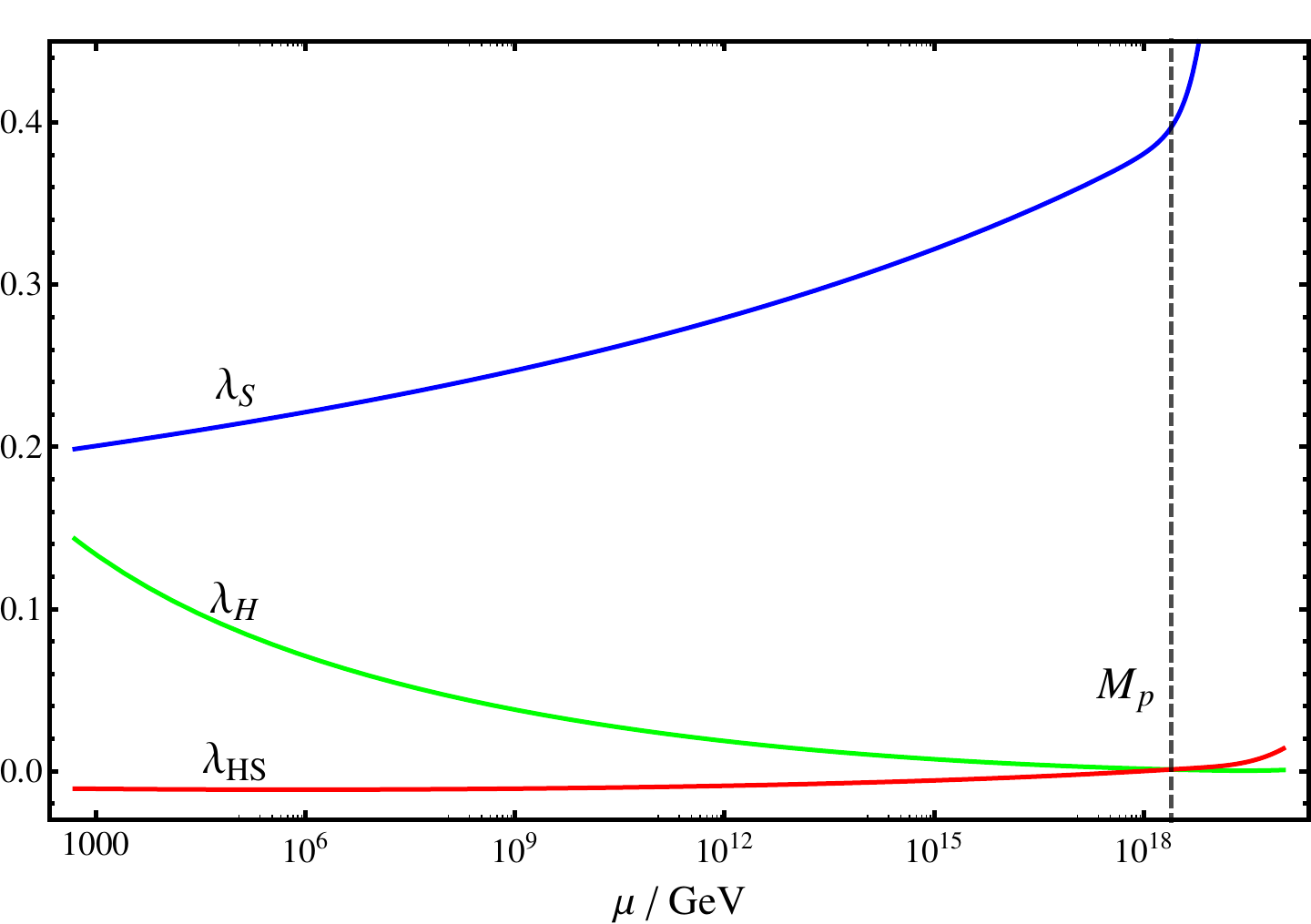}
\includegraphics[width=.48\textwidth]{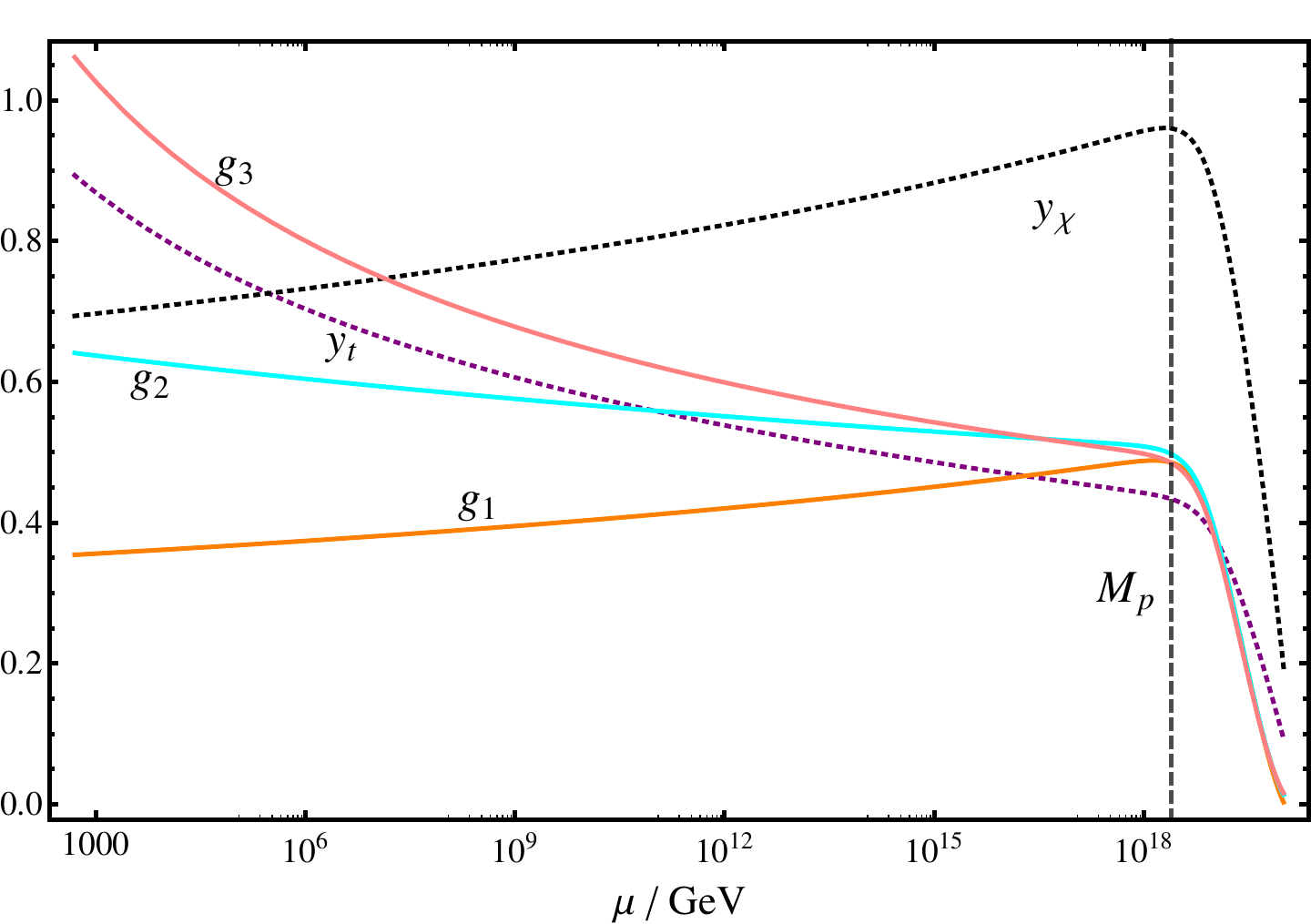}	
\caption{RG evolution of the couplings, where we fixed $Q_D = 1$ and $m_S = 500$~GeV, and used the initial conditions $\lambda_{H}(M_P) \approx 0$, $\lambda_{HS}(M_P) \approx 0$, $y_{\chi}(m_S)=0.69$ and 
$\lambda_S(m_S)=0.21$ as given in the text.
{The numerical values for the gra\-vi\-ty co\-efficient were taken universally to
be $\xi_0 = 0.024$, $a_\lambda = 1$, $a_y = -1$, and $a_g = - 1$. 
Large variations on these parameters have been investigated and no relevant differences were encountered.}.
}
\label{UVcomplete1}
\end{figure}

To investigate whether the asymptotically safe scenario agrees with the value of the discovered Higgs mass, we assume 
that $\lambda_H(M_P) \approx 0$ and $\beta_{\lambda_H}\big|_{M_P} = 0$, as prescribed in~\cite{Shaposhnikov:2009pv,Bezrukov:2012sa}. 
This effectively sets $\lambda_{HS}(M_P) \approx 0$.
The couplings $y_\chi$ and $\lambda_S$ are determined as in the previous section at the $m_S$ scale using Eq.~\eqref{y_chi-g1-ratio} and \eqref{lambda_S-y_chi-ratio}. We restrict the analysis of this section to $Q_D = 1$. This fully constraints the parameter space and leads to the evolution of the couplings shown in Fig.~\ref{UVcomplete1}.

The first thing to note is that $\lambda_S$ stays positive and perturbative all the way to the Planck scale as required by consistency of the asymptotically safe scenario.
The next thing to note is that $\lambda_{HS}$ stays very small (and negative) all the way down to the $m_S$ scale.
 This means that its effect on the running of $\lambda_H$ is negligible in the entire region from the Planck scale and down to the Fermi scale.
 Moreover, it does not ruin stability of the electroweak vacuum, since
 the potential is bounded from below as long as $2 \sqrt{\lambda_H \lambda_S} +  \lambda_{HS} >0$.
Thus the Higgs mass prediction from the pure SM within asymptotic safe scenario stays intact. We recall that the prediction is $m_H = 129\pm 6$~GeV \cite{Shaposhnikov:2009pv,Bezrukov:2012sa}.
In fact, the effect of $\lambda_{HS}$ is to push the Higgs mass prediction slightly down ($<1$~GeV).  Since  we have fixed the top mass to its experimental central value  and allowed the Higgs mass, at the electroweak scale, to be determined by the UV conditions above, the stability regions of Fig.~\ref{sexymoney}  will change slightly.

\begin{figure}[tb]
\includegraphics[width=.48\textwidth]{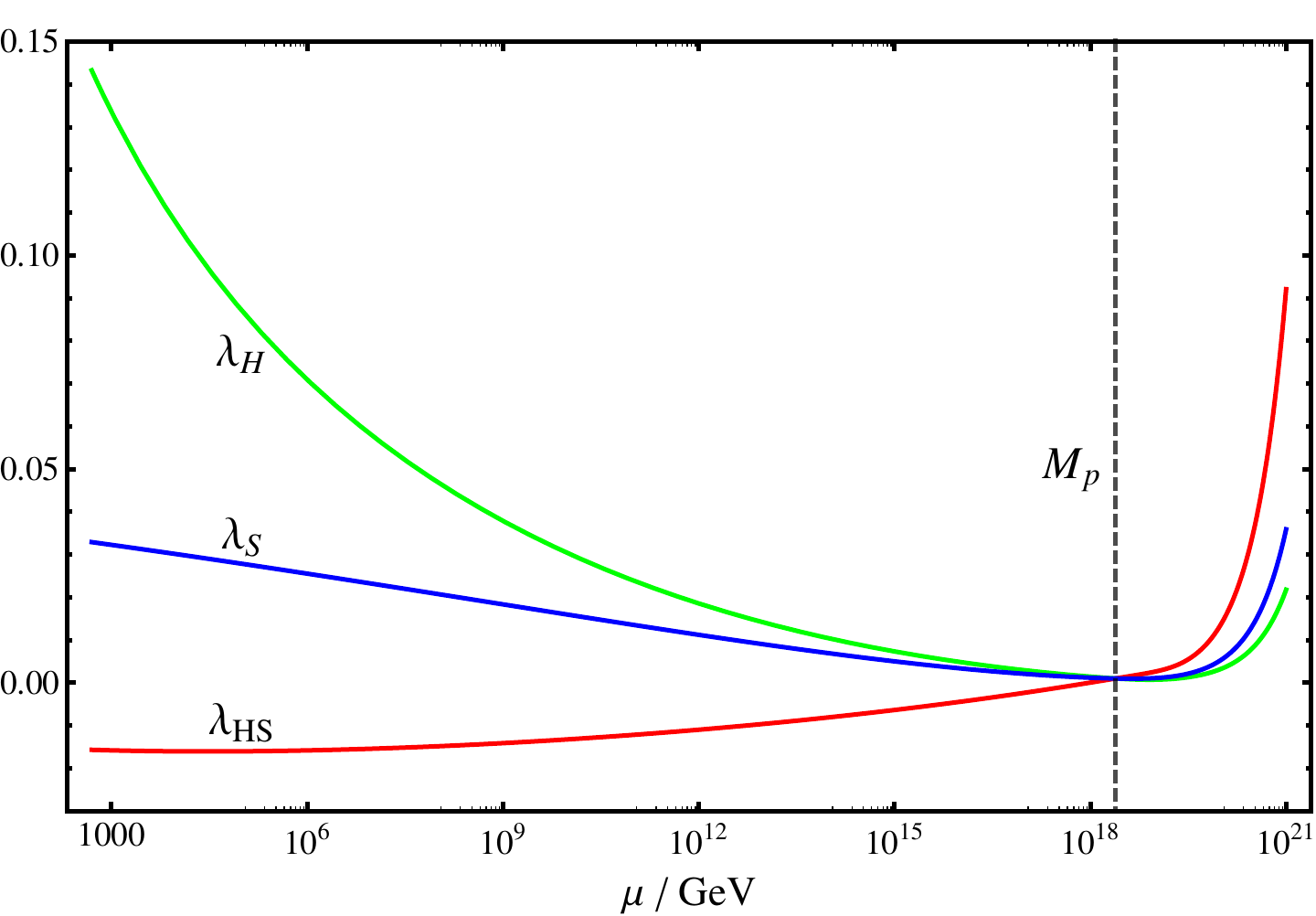}
\includegraphics[width=.48\textwidth]{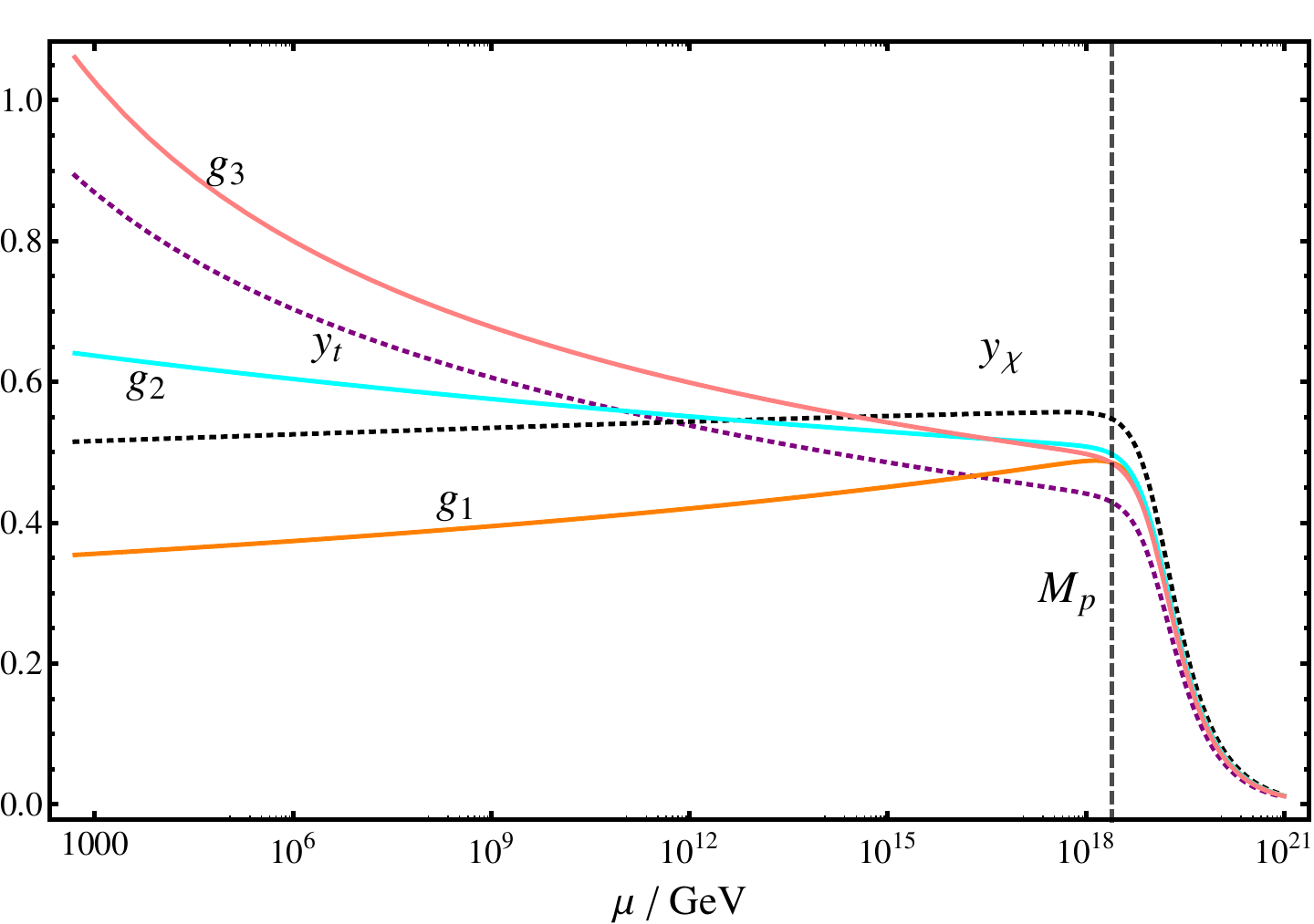}
\caption{RG evolution of the couplings, where we fixed $Q_D = 1$ and $m_S = 500$~GeV and used the Planck boundary conditions $\lambda_{H}(M_P) \approx 0$, $\lambda_{HS}(M_P) \approx 0$, $\lambda_S(M_P) \approx 0$ and $y_{\chi}(M_P)=0.54$, such that $\left.\beta_{\lambda_H}\right|_{M_P}\approx 0$ and $\left.\beta_{\lambda_S}\right|_{M_P}\approx 0$. 
{The numerical values for the gra\-vi\-ty co\-efficient were taken universally to
be $\xi_0 = 0.024$, $a_\lambda = 1$, $a_y = -1$, and $a_g = - 1$. Large variations on these parameters have been investigated and no relevant differences were encountered.}}
\label{UVcomplete2}
\vspace{-3mm}
\end{figure}

So far we have insisted in reducing the parameter space by using  the low energy boundary conditions coming from \eqref{y_chi-g1-ratio} and \eqref{lambda_S-y_chi-ratio} to determine $\lambda_S$  and $y_\chi$. 
One could argue, however, that a more consistent choice from the point of view of asymptotic safe gravity would be to require the vanishing of $\lambda_S$ and its beta function near the Planck scale, as done for $\lambda_H$. This corresponds to assuming $\lambda_S(M_P) \approx 0$ and $y_\chi^2(M_P) = \sqrt{\frac{3}{2}} Q_D^2 g_1^2(M_P)$ to ensure that $\beta_{\lambda_S}\big|_{M_P} = 0$. In this case, the prediction for the values $\lambda_S(m_S)$ and $y_\chi(m_S)$ changes to smaller values, while the effects on $\lambda_{HS}$ and thus $\lambda_H$ remains effectively unchanged. This scenario is shown in Fig.~\ref{UVcomplete2}, where again
the electroweak vacuum remains stable since
$2 \sqrt{\lambda_H \lambda_S} +  \lambda_{HS} >0$ along the
entire energy range (using very small, but positive values for the quartic couplings at the Planck scale).

Our study shows that the prediction of the Higgs mass from the interplay with asymptotic safe gravity, 
put forward in \cite{Shaposhnikov:2009pv,Bezrukov:2012sa}, apply to a wider class of extensions of the SM. These models generically contain new perturbative scalar and fermionic sectors. The key ingredients are to require, as done for the SM, that the Higgs self-coupling $\lambda_H$ and its beta function to be zero just below the Planck scale.
The vanishing of the beta function guarantees the absence of a Landau pole immediately above the Planck scale.
%
%
 We note that the Higgs mass prediction presented here, and in \cite{Shaposhnikov:2009pv,Bezrukov:2012sa} for the SM, are lower bounds compatible with the asymptotic safety scenario\footnote{If there were tree level threshold effects on the quartic Higgs self-coupling, like in Ref.~\cite{EliasMiro:2012ay}, the bound on the Higgs mass could be lowered further.}.
 
 An asymptotically safe scenario as the one depicted above, albeit being perfectly compatible with these kind of stable extensions of the SM, may be at odds with the further requirement to also feature a phenomenologically viable DM candidate. 
If it is assumed that $a_\lambda>0$ to ensure a highly predictive model, then the scalar couplings must vanish at the Planck scale.
 This assumption stems from certain quantum gravity computations. If, however, a negative sign is assumed this would  enable the combination of asymptotic safety and large scalar couplings around the Planck scale that can accommodate the correct DM thermal relic density at low energies\cite{Frandsen:2013bfa}.
 

\section{Conclusions}
\label{conclusions}

We studied 
the stability of the Higgs potential in $S\bar{E}\chi y$-like extensions of the SM.  We provided the first indication that, differently from the SM, the models can support a stable electroweak vacuum. This led to relevant constraints on the parameter space of the extended sector.

 We next added the gravitational corrections to the beta functions of the theory within an asymptotic safe gravity scenario.  This framework requires the scalar couplings and their beta functions to be near vanishing at the gravity transition scale. We showed that it is possible to satisfy these UV conditions at the price of making the theory less compatible with a stable electroweak vacuum.  
 
 Our analysis shows that the $S\bar{E}\chi y$-like extensions of the SM, which provide dark matter candidates, may at the same time resolve the metastability issue in the Higgs potential. This adds to the favourable features of these extensions presented in \cite{Dissauer:2012xa,Frandsen:2013bfa}.
 Under the assumptions for the sign of the coefficients $a_i$  we have made in this work for the asymptotically safe gravity scenario, the investigated models cannot accommodate a single DM candidate as thermal relic unless further extended, or the original assumptions changed.

  
 In the near future we envision to improve and generalize on this exploratory analysis by including higher order corrections following the mathematically consistent way to perform the perturbative analysis following \cite{Antipin:2013sga,Antipin:2013pya}.

\acknowledgments
This work is partially supported by the Danish National Research Foundation DNRF:90 grant.

\end{document}